ALESSIO EMANUELE BIONDO, ALESSANDRO PLUCHINO, ANDREA RAPISARDA (2012)

# RETURN MIGRATION AFTER BRAIN DRAIN: A SIMULATION APPROACH

 **Abstract**

The Brain Drain phenomenon is particularly heterogeneous and is characterized by peculiar specifications. It influences the *economic fundamentals* of both the country of origin and the host one in terms of human capital accumulation. Here, the brain drain is considered from a microeconomic perspective: more precisely we focus on the individual rational decision to return, referring it to the social capital owned by the worker. The presented model compares utility levels to justify agent's migration conduct and to simulate several scenarios within a computational environment. In particular, we developed a simulation framework based on two fundamental individual features, i.e. risk aversion and initial expectation, which characterize the dynamics of different agents according to the evolution of their social contacts. Our main result is that, according to the value of risk aversion and initial expectation, the probability of return migration depends on their ratio, with a certain degree of approximation: when risk aversion is much bigger than the initial expectation, the probability of returns is maximal, while, in the opposite case, the probability for the agents to remain abroad is very high. In between, when the two values are comparable, it does exist a broad intertwined region where it is very difficult to draw any analytical forecast.

 **Introduction**

The rationale behind brain drain migration is complex and heterogeneous. More than an introductory sentence, this rather seems a conclusion and we are aware of it. Nonetheless, since in order to analyze any phenomenon an authentic and preliminary recognition is needed, we better start just from the beginning. The very first question is: 'What is *brain drain*?'

Kwok and Leland (1982) say, (p.91): «the "brain drain" is an expression of British origin used to describe […] skilled professionals who leave their native lands in order to seek more promising opportunities elsewhere». From a home economics point of view, one should argue then, brain drain exists only in case of high-skilled emigration, when relatively high qualified workers decide to move from their own country to a foreign one, since they are relatively over skilled and (nonetheless) they cannot obtain a job position that fits their advanced skills at home. Truly, it is worth noticing that they have an option: by accepting either less qualified or less lucrative job positions in their own domestic labour market. In other cases, generally speaking, emigration occurs when workers are induced, by adverse economic conditions in their domestic economy, to migrate looking for *any* job opportunities abroad. Differently from the high skilled emigrants, low skilled emigrants have no valuable options in their own country and their movements are based on, let us say it roughly, survival reasons. The migration caused by poverty, unemployment, rigid domestic labour markets, is a phenomenon that reveal the hope, the *individual chance* to find an income, in order to escape from poverty trap.

Brain drain migration is completely different and characterizes the circumstance of a worker who invested in terms of human capital on himself/herself and wants the consistent return. The second question, then, follows in turn: 'Who are *brains*?'
They are skilled workers, as we said above. It means that no matter which sector they are involved in, they are the highest kind of workers that their country can generate, by definition. Their qualification is very high, their native country cannot hire them all, and thus a share of this high-skilled cohort of workers is forced to leave, unless they accept less prestigious, less lucrative positions in order to remain at home. One quite common example is the case of academic personnel, such as researchers, lecturers, and Ph.D. students who have not structured placement in Universities of their native country, and therefore have to decide to migrate to look for career and success abroad. In fact, this is the approximate trade-off that *brains* face: "should I remain at home, with my family, my friends, in my city where I belong, with a job position that does not allow me to use my (accumulated) human capital, that gives to me a lower salary and limited perspectives *or* should I leave, going abroad, looking for 'my chance', following my educational standard, challenging myself and obtaining *in time* what I hardly studied for?"

Said this way, the reader can understand that there is not a unique solution, because every person has his/hers own. And, of course, it is linked to the cost of the accumulation of the human capital: in such a rationale, the brain drain migration is a sort of *extreme defence* to ensure adequate revenue to invested resources in human capital development. The decision to migrate is, of course, never simple and, furthermore, it is coupled to the complementary choice about whether to return.
Once the brain drain migration occurred (we are therefore referring exclusively to those who decided to leave their own homeland seeking for the *right* job, adequate to their skills), the second problem is about the permanence in the host country. Of course, it can be either temporary or permanent. Thus, who once in his/her life decided to migrate, has to solve the second puzzle: "should I return home, or should I remain here?" As in the first case, this is not a simple problem, since many aspects are, again, involved. Career addiction and adequate rewards to human capital investments are strong reasons to remain abroad, but love and affection for family or friends, environment and many other life standards-related reasons are strong as well.

How is the final decision taken? Can we imagine that it works like a balance? How strong is career addiction? Does the time spent abroad affect the final decision? How? What if career advances and partner/sons are *not* on the same side? Questions like these are at the core of the microeconomic approach of the brain drain phenomenon. This perspective is focused and developed on an individual point of view, somehow related to individual preferences and wage differentials. Among many others, such aspects can be read in: Glystos (1988), Dustmann (1994) and (1995), Borjas and Bratsberg (1996), Bell (1997), Friedberg (2000), Dustman and Weiss (2007), Mayr and Peri (2008), Biondo *et. al.* (2012). Data and theoretical models have often been used in order to solve the return puzzle. The reason of this relevant attention is that the return decision affects the economic dynamics of a country. This is the second approach to the brain drain phenomenon: the macroeconomic one. The decision to migrate abroad impoverishes the source country, since the best human capital is subtracted and given to the host country.

Exactly like a football match where the score is economic growth, this becomes a strong advantage for production and income distribution: where most productive and qualified workers are, there are development, high living standards, economic growth, good public services, and so on. Thus, the country of origin remains behind and becomes always less attractive for future brains coming in future generations. Conversely, if after a period spent abroad, *brains* decide to come back home, they can improve the average level of productivity of population, with their accumulated human capital. As it appears clearly, this means that at a macro level, temporary brain drain can be an opportunity. To these macro-consequences of brain drain is dedicated a very large part of literature, such as Beine, Docquier, and Rapoport (2001), Commander, Kangasniemi and Winters (2003), and Beine, Defoort, and Docquier (2006), Lucas (1988), Azariadis and Drazen (1990), Haque and Kim (1995), Lucas (2004), Cox, Edwards and Ureta (2003), Poutvara (2004), just to mention a few.

Our work is focused on the microeconomic dimension of the brain drain migration. We conjecture that the worker has already migrated and we study what happens to his/her life and how time spent abroad can influence the rational process of determination of the final choice in terms of return or definitive permanence abroad. In order to do so, we try to compare two opposite components of this individual choice to find the optimal decision.

Sometimes, the need of data leads to shape too directly the profile of the investigation on the characteristics of a given country. This is the consequence of direct analyses of country data (when available), based on the empirical interview of subjects seeking reasons to leave or to return. Since we think that a broader analysis is to be done before fitting any dataset, our approach will not refer to any specific country, trying to obtain a general framework, able to describe the dynamics of the return migration after brain drain. The reason underneath this decision is that at country-level, reasons for brain drain migration and opportunities for return flows are almost everywhere somehow similar and approximately linked to aggregate expenditure in research, opportunities for funding, merit weight in career progression, and so on. In a forthcoming article we plan to consider a real database (at the moment under construction) to calibrate model's parameters in order to get some insight about individual profiles.

We present our study in two phases. In the first one we consider an analytical approach and show the dynamics of the decisions concerning out-flow brain drain and in-flow return. In the second one we develop several simulations, which will allow us to consider in detail the various components of the individual return choice. More precisely, in section two we present the theoretical framework of our model, in section three we show the details of the computational model and discuss the simulation results. Finally, in section four we will draw some conclusions and discuss future developments.

##  The Rational Mechanics of Migration Decisions

In order to build our framework we need to follow a simple logic path that describes the rationale behind a migration decision. The present value of a variable that encompasses most relevant aspects of life can be suitable. Economic models often refer to the utility function that associates individual happiness with consumption. We here adopt a broader concept than consumption, since we want to underline many other aspects of individual's life that are linked with social interaction. We name this variable 'social capital' and it is intended to include, e.g., career or wage advancements, job stability, life conditions, and so on. It is worth to notice that the concept of social capital has been widely discussed in economic and sociologic literature, and of course we do not aim to participate to this debate. However, we consider the social capital as an individual resource, whereas other existing definitions are referred to different aspects at a social level (as reviewed in Paldam 2000 and Portes 1998). The advantage of our methodological choice is that we can focus on social interaction from an individual point of view, more than simply on the consumption of a basket of goods: thus, we can characterize spatial and time-specific elements. Eventually, consumption can be reasonably enjoyed everywhere, whereas social and personal interactions among people and their consequences on individual choices are specific with respect to time and location.

Of course, there exists the theoretical possibility to model differences in utility given by same consumption levels according to the individual location (whether at home or abroad): this approach has been recently presented by Biondo (2012) and leads to the determination of the optimal timing of the return decision. However, the simulation framework gives the chance to focus on many dynamical aspects that, although coherent at a macro level with the theoretical investigation, are very difficult to be completely captured with an exclusively analytical approach (see e.g., Silveira J.J. *et al.* 2006, Garcìa-Dìaz C. and Moreno-Monroy A.I., 2012). For such a reason, the approach of Biondo (2012) has been widened here by defining social capital as a composite value that melts together the personal value given by each individual to the three main components (job, family, and friends) of his/her contacts, both at home and abroad. As time passes, the migrant dynamically develops a net number of personal links with other individuals in these three groups, thus creating his/her personal contacts network abroad (a very simple network with a 'star' topology), while a similar network already exists at home. Since the migrant's welfare (let us name it, simply, utility) depends on these two contacts networks, the choice of a location to live in is somehow based on them, too. As it appears immediately, we are trying to show how life can be "evaluated" in migrant's mind in a comparative process between life at home and abroad.

Since we are looking exclusively to *brains*, we implicitly never forget that they can count on an option wage in their home country, coming from the low-profile job choice. The approach followed here will not therefore optimise any function subject to any budget constraint in order to find an optimal migration decision. On the contrary, we will analyse the comparison between *prospective* life in the home country and *prospective* life abroad, period by period, so that the final choice can be simulated as a result of the comparison between levels of expected utility in both locations. The income constraint will not be represented as a bounding function; it will, instead, participate to utility generation by career development, in the sense of greater opportunities to develop further social capital. Doing this simplification we can simply use the social capital level as a *proxy* of income.

We assume that time runs from $t = 0$ (the initial moment of the out-flow brain drain migration) to the final moment $t = T$, when the migrant chooses no longer to participate in the labour force. For simplicity, since we do not focus on present value of pensions, we will not deal with retirement issues: in order to do it, the last event we care of is the exit from the labour force (the reader can eventually assume that $t = T$ is the moment of death). Thus, our model will look for the existence of the moment $t = \sigma$, in the interval $]0,T]$, when the worker possibly decides to return.

At the moment of migration ($t = 0$) the worker has a given amount of social capital in his/her country, resulting from the summation of all his/her relational directed links, established with parents and family in general, with friends, and with his/her job environment. We define the social capital $c^H(t)$ at home as:

$$c^H(t) = \sum_{i=1}^{N^H(t)} w_i^H R_i^H \qquad (1)$$

where $w_i^H > 0$ are the weights of the links that the worker has in home country with his/her $N^H(t)$ personal contacts and $R_i^H$ (included in the interval $]0,1]$) is the relevance of each contact (the differentiation of links will be detailed later in the simulation section). Notice that the number of contacts $N^H$ does depend on time: at $t = 0$ it has its maximum value $N^H(0)$, then it starts to decrease for $t > 0$ since after migration it is reasonable to presume (as we will do) that job links may fade in time.

Similarly, the social capital in the foreign country $c^F(t)$ can be defined as

$$c^F(t) = \sum_{j=1}^{N^F(t)} w_j^F R_j^F \qquad (2)$$

where $w_j^F > 0$ and $R_j^F$ are intended as foreign counterparts of aforementioned $w_i^H$ and $R_i^H$, respectively, while $N^F(t)$ is the number of foreign contacts (in family, friends and job groups). In what follows, we will assume that at $t = 0$ the migrant has always only one contact abroad, i.e. $N^F(0)=1$, which is an element of his/her foreign job group whose relevance is $R_1^F$. This implies that it will be $c^F(0)=w_1^F R_1^F$ : this value is a minimum, since the foreign social capital is intended to increase in time as long as the number of contacts abroad increases.

The utility, either at home or abroad, is a function of social capital in any moment of time,

$$u(t) = u\left[c^k(t)\right] \qquad\qquad k = H, F \qquad (3)$$

Following the usual assumptions done about the utility function in economic literature, we will consider that here utility is a monotonic non-decreasing function of the social capital: thus

$$\frac{du[c^k(t)]}{dc^k(t)} > 0 \quad \underline{\text{and}} \quad \frac{d^2u[c^k(t)]}{dc^k(t)^2} < 0 \qquad , \qquad k = H, F \qquad (3.a)$$

The utility function that will be adopted here belongs to the constant absolute risk aversion (CARA) family. The reason inspiring such a choice is that we do not take in account any change in the risk aversion of the subject due to variations in the owned social capital: we choose randomly the level of risk aversion $a$ (in the interval $[0,1]$) at $t = 0$ for any person and it will remain constant through the entire time lag. Thus we have

$$u[c^k(t)] = 1 - e^{-ac^k(t)} \qquad k = H, F \ ; \ u \in [0,1[ \qquad (3.b)$$

In our model we assume as given the decision to migrate: therefore at $t = 0$ we surely have $c^H(0) \gg c^F(0)$, which in turn implies $u[c^H(0)] \gg u[c^F(0)]$. In order to justify the brain drain out-flow migration decision we introduce a positive expectation $E(0)$ such that, added to $u[c^F(0)]$, it at least overcompensates the domestic utility $u[c^H(0)]$. Therefore, the condition (always satisfied by definition) for migration at $t=0$ results to be

$$u[c^H(0)] < u[c^F(0)] + E(0) \qquad (4)$$

where $E(0) = u[c^H(0)] + R_1^F$ is the initial expectation term, equal to the sum of the domestic utility at $t = 0$ plus the relevance of the first (unique) initial contact abroad. The expectation represents the idea (and the hope) that the migrant has about his/her capacity to meet new people, create new social links in order to build abroad the same kind of contacts networks (with 'star' topology) he/she had at home (job, friends, and possibly family, in terms of partner, sons, etc…). The rationale behind this assumption is that in order to leave, the worker must evaluate his/her utility abroad at least as much as his/her utility at home, added by a rough indicator of the probability of success abroad, which is the relevance of his/her first contact abroad.

After the brain drain outflow migration ( $t > 0$ ), the expectation decreases naturally in time at a given rate $r_E(a)$ that we assume directly proportional to the individual risk aversion coefficient $a$:

$$E(t) = E(0) - r_E(a)t \qquad \underline{\text{with}} \ r_E(a) \propto a \qquad (5)$$

i.e. the expectation decreases very quickly for highly risk adverse subjects and *vice-versa*. It is worth to notice that in the term "risk aversion" we include all those individual features (patience, competence, self-confidence, etc.) able to influence the expectation decay. On the other hand, foreign social capital $c^F(t)$ - and therefore $u[c^F(t)]$ - tends to increase with the number of new contacts abroad. Thus, for $t > 0$, the condition for the worker to remain abroad becomes:

$$u[c^H(t)] < u[c^F(t)] + E(t) \qquad (6)$$

Let us recall that the social capital at home $c^H(t)$, starting from its maximum value $c^H(0)$, is assumed to decrease in time due to vanishing job contacts (and, consistently, the same dynamics will affect the utility function $u[c^H(t)]$). Actually, we also introduce the possibility that the Government in the home country could apply policies to *call* brains back. We implement such an opportunity in the model by allowing the utility at home to rise again to its maximum value, $u[c^H(0)]$, for the limited validity period of the call.

Summing up all these considerations, if at any moment in the time interval $]0,T]$, say $t = \sigma$, should the inequality (6) revert its direction (or even in case of equality), the worker would return to his/her native country. This happens if, further substituting eq. (5) in eq. (6), it holds

$$u[c^H(\sigma)] \geq u[c^F(\sigma)] + E(0) - r_E(a)\sigma \qquad (7)$$

i.e. if the growth of the foreign utility (function of the social capital accumulated during time spent abroad) is no more sufficient to compensate the decreasing path of expectation in order to overcome the utility at home. Assuming equality in eq.(7), one could straightforwardly derive the following expression for the return moment

$$\sigma = \frac{u[c^F(\sigma)] - u[c^H(\sigma)] + E(0)}{r_E(a)} \qquad (8)$$

which somehow looks like the main finding of Biondo (2012) obtained through an entirely analytical approach.

On the other hand, it is absolutely non trivial to predict if the condition (7) will be satisfied for some $\sigma < T$ since, for a given individual, it strongly depends on the combination of many parameters: the initial configuration and the dynamical time evolution of the contacts networks (at home and abroad), the initial expectation and its decay rate, and the individual risk aversion. For this reason, as previously anticipated, we will adopt here a simulative approach in order to get statistically relevant insights on the return migration occurrence. As we shall see later, such an approach has also the advantage to be more suitable for a direct comparison with real data.

The reader should be also aware that the professional characterization of the migrant's job is very important. In our paper, we often refer to the academic brain drain phenomenon since we focus on it primarily, but one could argue that the timing of the return (if any) is different for different kinds of workers: even in identical conditions, different jobs can suggest different return decisions.

In the next section we describe the implementation of our model within a computational environment and we present the main results of the simulations.

 **The Computational Model: Description and Simulation Results**

Before starting with the simulations, a couple of examples may fruitfully amend the given theoretical scenarios. Consider the case of a person who is not able to develop sufficiently his/her social capital. Many reasons can explain this profile: scarce propensity to social contact, scarce personal ability and/or insufficient skills, low propensity to social integration, misfortune, and so on. In all of these cases $c^F(t)$ will remain low, and it will not grow "sufficiently" in time: this intuitively leads to revert the direction of inequality (7) and to return to home country. Similarly, an excess of personal risk aversion can erode too drastically the expectation value, obtaining again the returning migration result.

On the contrary, consider the opposite case of a very dynamic person, socially appealing and able to integrate himself/herself within the new community, maybe lucky to find opportunities or being in the right place at the right moment, or maybe so greatly skilled to be highly productive and therefore particularly appreciated in his/her new job context. In these cases the growth in $c^F(t)$ will be so high to overcompensate the decrease in expectation in such a way that inequality (7) holds true, even forever. Another example should be done referring the aforementioned possibility that the Government in the home country applies a policy to call brains back. In this case, the effect on our model will appear as a boost on the domestic social capital value that increases the utility level at home, inducing more probability for a return. In the following subsections we will try to embed all these features in a prototypical computer model of brain drain migration in order to obtain reliable predictions about this complex phenomenon.

**Contacts Network and Social Capital at t =0**

Let us start by describing the contacts network the migrant has developed at home until the moment of brain drain out-flow migration. As previously seen, it is made of three main components (job, family, and friends), from which we will be able to calculate the social capital at home at $t = 0$ by means of eq.(1).

In Figure 1 we show an example of such a network, where the migrant (agent in red) is placed at the centre of his/her world, among the three components of his/her own contacts. As a consequence, the resulting topology has a simple 'star' shape with directed links, since the information flows only towards the migrant. In the upper part there is the hierarchical job colleagues' pyramid, whereas in the lower part the reader can find the family group (bottom-left) and the friends group (bottom-right). We indicate with the term "network" also the job and friends groups (see labels in the figure) since of course, in reality, these agents are also variously linked to each other, even if in our model we explicitly consider only their ties with the migrant. We colour in yellow the agents linked with the migrant, at a given time, and in dark yellow the others.

At *t = 0* a given configuration of links is randomly generated with uniform distribution: in this case, we see that the migrant has two children and one brother/sister, seventeen friends, but no partner and no parents. In our model, at each link is assigned a weight $w_i^H > 0$. These weights are identical for all the members of each group, but their values are different among groups. In particular, the weights are proportional to the intensity of the relationships; therefore, we assign the maximum value ($w_i^H = 10$) to the family links, followed by job links ($w_i^H = 5$) and, finally, by friends' ones ($w_i^H = 3$).

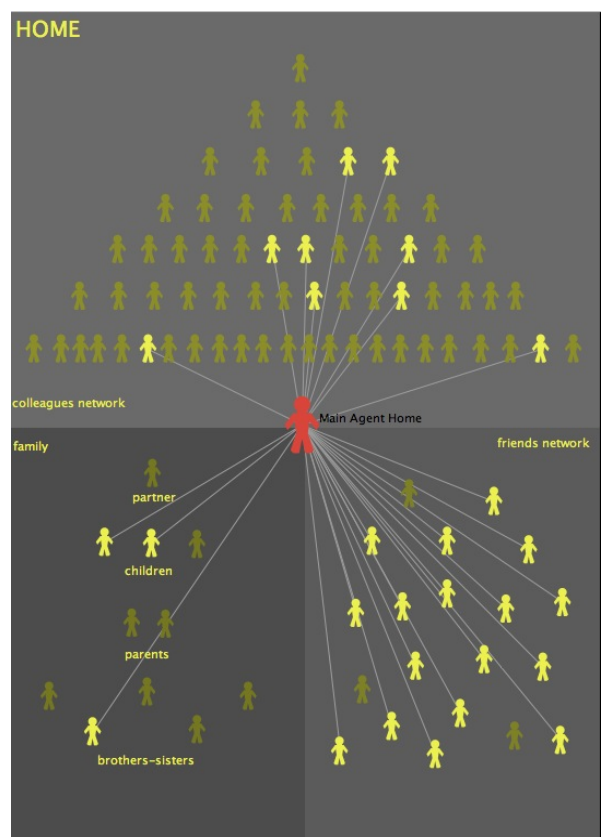

**Figure 1:** Initial social capital network at home

**Table 1**: Settings for the relevance values at home and abroad ( $k = H, F$ )

| FAMILY | JOB COLLEAGUES | FRIENDS |
|---|---|---|
| parents: $\quad R_i^k = 1.0$ <br> children: $\quad R_i^k = 1.0$ <br> partner: $\quad R_i^k = 0.8$ <br> brothers-sisters: $\quad R_i^k = 0.5$ | Randomly selected in the following intervals (with uniform distribution): <br> level-1: $0.0 \le R_i^k < 0.1$ (bottom) <br> level-2: $0.1 \le R_i^k < 0.2$ <br> level-3: $0.2 \le R_i^k < 0.4$ <br> level-4: $0.4 \le R_i^k < 0.6$ <br> level-5: $0.6 \le R_i^k < 0.8$ <br> level-6: $0.8 \le R_i^k < 1.0$ <br> level-7: $\quad R_i^k = 1.0$ (top) | Randomly selected in the interval $R_i^k \in\ ]0,1]$ with uniform distribution |

At the same time, we also differentiated the members of the three groups according to their relevance $R_i^{\,H}$ (included in the interval $]0,1]$), which is the different importance that each member plays in migrant's life. We set the relevance values as described in Table 1. By using these values for the weights of the links and the relevance of the agents, for the random configuration shown in Figure 1, it is possible to calculate the initial social capital at home $c^H(0)$ with eq.(1) and the corresponding initial utility $u[c^H(0)]$ by means of eq.(3.b). Consider that in the latter question, we had to rescale the values of social capital (multiplying it by a scaling factor equal to 0.02) in order to let the utility go to one when the social capital is maximum. This is not a loss of generality since it is equivalent to a rescaling of weights of eqs.(1) and (2), which are arbitrarily defined, as just described above.

Let us now describe the initial configuration of the contacts network abroad. As previously said, we assume that at $t = 0$ our migrant, represented by the usual red central agent, has only one contact abroad: in particular, a job contact whose relevance $R_1^{\,F}$ will influence his/her expectation $E(0)$. We reasonably assume that the hierarchical level of such initial job contact abroad cannot be higher than the highest job contact own at home. For example, since in Figure 1 the most relevant job contacts at home are at level-5, this means that the position of the initial job contact abroad will be randomly chosen among the first five levels (from the bottom).

In Figure 2 we show an example of such a random initial configuration, where the initial job contact (in blue) is located at the second hierarchical level and is linked to the migrant with a given weight $w_1^{\,F}$ (the other no-linked contacts are represented in light blue). In order to assign the random values to both the weights and the relevance abroad we will follow the same rules adopted in the home case (see again Table 1 for the relevance-values intervals).

We also assume that the migrant starts his/her career abroad at the bottom level of the hierarchical pyramid, as indicated by the little replicated red agent (avatar) on the left. As we will see, the latter will have the possibility to climb the hierarchy at $t > 0$ depending on the credits he/she will earn by linking new job contacts during the time spent abroad. These credits at a certain time are simply the sum of the relevance of all the job contacts existing abroad at that time: periodically, according to both the credits amount and the risk aversion, the migrant has the possibility to be promoted to the next level in the hierarchy (with consequent upward shift of his/her little avatar in Figure 2). Of course, meanwhile, the migrant could develop new links also with the friends and family components in order to increase his/her foreign social capital, calculated by eq.(2), and his/her utility abroad, eq.(3.b). At the end, also the resulting network abroad will assume a 'star' topology, with the migrant as central node.

Once defined the two initial networks, at home and abroad, and calculated the respective initial social capital and utility levels, before starting the simulation we have to assign the risk aversion $a$, randomly chosen in the interval $[0,1]$ with a uniform distribution, and to calculate the initial expectation by the given formula $E(0) = u[c^H(0)] + R_1^F$. Actually, in our simulations these two parameters completely characterize the initial condition of a given migrant; therefore we will focus on them in order to classify the return time distribution over many different configurations of initial conditions (events).

**Single event dynamics of return migration**

In this subsection we present the dynamical evolution of single events of return migration assuming a total time interval $T$ of 15 years. We adopt one month as discrete time unit, since we consider such a time scale the best compromise in order to take into account both migrant's short-term decisions (career advances and links formation) and his/her long-term behaviour (return or not).

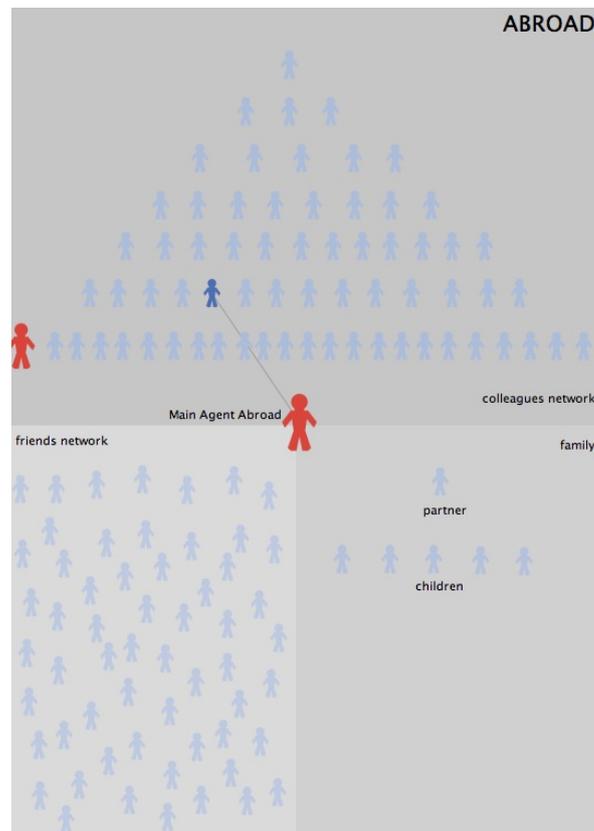

**Figure 2:** Initial social capital network abroad

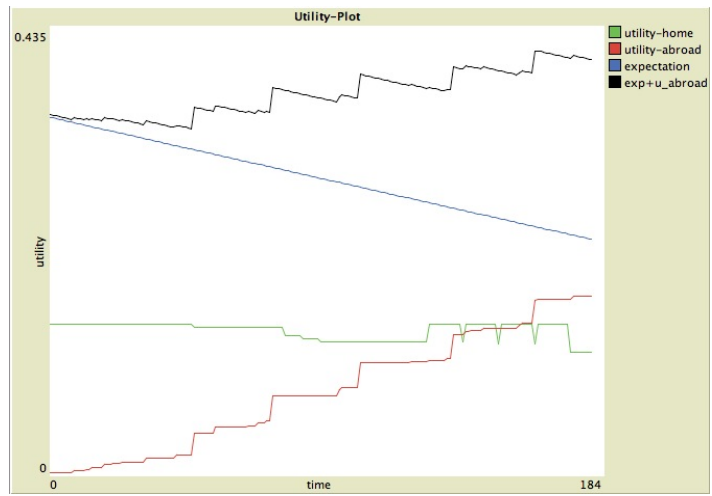

**Figure 3:** Single event dynamics of return migration (no return case)

The first event we consider starts from the initial configurations (at *t = 0*) shown above in Figure 1 and Figure 2. In this case the relevant initial parameters are the following:

*$c^H(0)$=60.5, $u[c^H(0)]$=0.146, $c^F(0)$=1.0, $u[c^F(0)]$=0.003, $R_1^F$=0.2, a=0.13, E(0)=0.346*

As one can see, the condition for brain drain out-flow migration at $t = 0$ stated in eq.(4) is satisfied, since the sum of the expectation and the utility abroad (0.349) overcompensates the utility at home. Running our simulation we want to check whether, for some *t > 0*, the latter inequality would be reversed: in other words, we look for a value *t = σ*, in the interval ]*0,T*], which would satisfy the return condition stated in eq.(7). In Figure 3 we show simultaneously, as function of time: the utility at home $u[c^H(t)]$ (green line), the utility abroad $u[c^F(t)]$ (red line), the expectation $E(t)$ (blue line) and the sum $u[c^F(t)]$ + $E(t)$ (black line). We recall that, according to eq. (5), expectation $E(t)$ decreases in time with a rate proportional to the risk aversion. Now, we assume that $r_E(a) = k_E a$, with $k_E = 0.005$, that will remain fixed for all our simulations (we will return on this issue later).

Firstly, we see that the utility at home decreases quite slowly since its decrement is due only to the progressive, random, deletion of home job contacts with a probability linearly increasing in time (simulating an understandable fading of relationships). Notice that, after about 10 years, our migrant received a sequence of four calls by his/her academic institution at home (corresponding to the case of adequate corresponding policies). As previously explained, in correspondence of each call, the utility $u[c^H(t)]$ is raised up to its initial value $u[c^H(0)]$ for a twelve months time lag, but in this case with no effects. In fact, even if the expectation abroad $E(t)$ progressively decreases in time, on the other hand the increment in foreign social capital, generated by the creation of new links with the contacts abroad (job, family and friends), visibly pushes upward the foreign utility $u[c^F(t)]$ such that the sum $u[c^F(t)]$ + $E(t)$ never goes below the home utility. Therefore, eq.(7) is never satisfied and our migrant will remain abroad permanently.

It is worthwhile to look at the final configuration of both the home and foreign networks in order to further explain the behaviour observed in the previous figure. In Figure 4 we report the status of the two networks at 180 months (15 years) after migration. We immediately recognize that, while the family and friends components at home (left panel) are unchanged with respect to the initial configuration of Figure 1, the number of the home job contacts has been reduced of six units (agents in red, no more linked to the migrant): this justifies the slight decrease of the social capital at home.

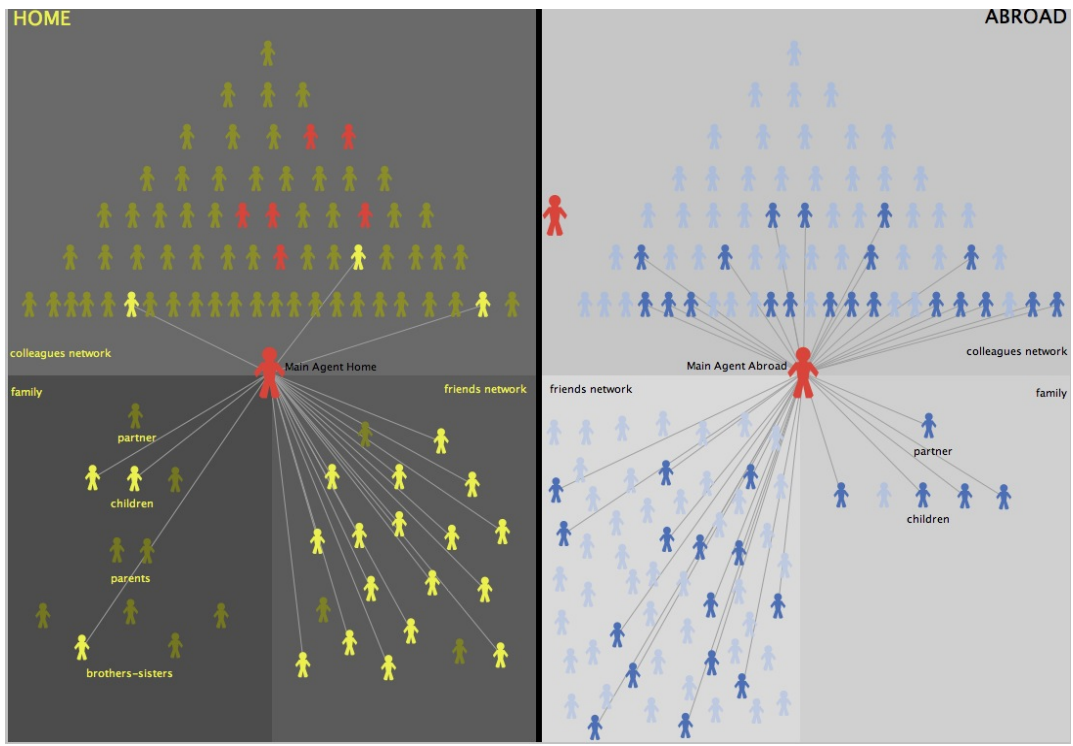

**Figure 4:** Social capital networks after 15 years (no return case)

On the contrary, the situation abroad (right panel) is much more complex than that shown in Figure 2: in fact, after migration, our worker rapidly developed new relationships in the foreign country that in turn raised his/her social capital abroad. In particular, we see that the migrant: a) has improved his/her position in the job hierarchy, gaining two promotions and therefore reaching the third level; b) has built a relevant group of friends; c) has found a partner abroad, with whom raised a family with four children.

Let us deepen these points.

First, in our model, at a given time, calling level-*m* the migrant's actual level and level-*z* that one of his/her initial contact abroad, the migrant is able to create job links only with workers at a hierarchical position *lower or equal* to level-*n*, with *n=m+z*.

Second, we assume that the addition of a new friend abroad depends on two parameters:
- the *probability $p_1(t)$* that the migrant meets a potential new friend (among light blue agents); initially we set $p_1(0) = (1-a)$ , being *a* the risk aversion;
- the *probability $p_2$* that, once the potential new friend has been met, he/she actually becomes a new friend; such a probability depends on the number of links (connectivity) of the potential friend in his/her latent social network and its value does not vary in time but is randomly assigned (with a uniform distribution) as initial condition. As time passes, new friends (blue agents) join the migrant's network and this affects the value of $p_1(t)$, which becomes equal to their average connectivity at time *t*. This means that, the more connected migrant's friends, the higher the probability for the migrant to meet new potential friends.

Finally, the *probability* to find a partner abroad is given by the product $p_1(t) \times (1-a)$; therefore, it is both inversely proportional to the risk aversion and directly proportional to the average connectivity of the migrant's friends; notice that, once a partner has been found, the probability to have children abroad becomes positive.

Summing up, comparing the two contacts configuration, at home and abroad, shown in Figure 4, it is not difficult to explain the final decision of no return, especially since the risk aversion of our migrant was quite small ($a = 0.13$) while the initial expectation was sufficiently high ($E(0) = 0.346$) with respect to the social capital at home ($u[c^H(0)]=0.146$). On the other hand, the decision to remain abroad has been strictly influenced by the unpredictable development of the social contacts in the three components of the foreign network: another migrant, also starting from identical initial conditions, might develop a completely different configuration for his/her social capital abroad and consequently reach a different final decision (as we will show later).

Let us now illustrate another single migration event but with an opposite conclusion.
The new initial configuration both at home and abroad is shown in Figure 5. In this case the relevant initial parameters are the following:

*$c^H(0)=95.7$, $u[c^H(0)]=0.501$, $c^F(0)=0.5$, $u[c^F(0)]=0.004$, $R_1{}^F=0.1$, $a=0.36$, $E(0)=0.601$*

Again, the condition for brain drain out-flow migration at $t = 0$ is satisfied, but the contacts configurations are quite different than in the previous simulation. First, in this case the family at home is much bigger than before: apart from two children, already present also in Figure 1, both parents are now still alive, the partner exists together with three more brothers-sisters. This new family configuration is sufficient to explain the increment of social capital at home with respect to the previous case: in fact, it overcompensates the poorer job contacts configuration (made by just two colleagues), while the friends component is actually unchanged.

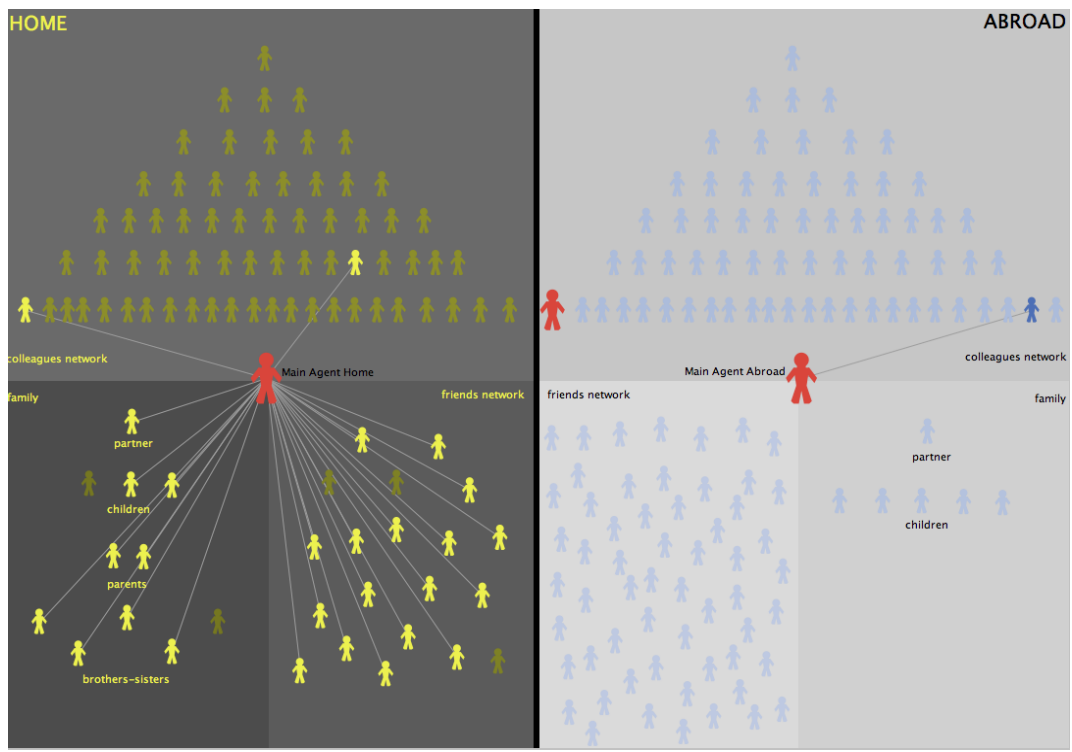

**Figure 5:** Social capital networks after 15 years (return case)

Concerning the initial contact abroad, it is now at the bottom level of the hierarchy, whereas in the previous simulation it was a little bit more relevant. Finally, it is worth noticing that our two key-variables, i.e. the risk aversion and the initial expectation, are both higher in value: this, on one hand, should increase the probability for the migrant to return at home (due to the higher risk aversion) but, on the other hand, should also increase the probability to stay abroad (due to the higher expectation). Therefore: what will the migrant decide? Our simulation will answer the question.

As Figure 6 shows, in this case, a value $t = \sigma$ (in the interval $]0,T]$ ), which satisfies the return condition in eq.(7) does exist: in fact, after exactly 95 months, i.e. almost 8 years, the sum of the expectation and the utility abroad is no longer sufficient to hold the migrant in the foreign country (where the black line crosses the green one). Such a result can be understood by looking at the final configurations shown in Figure 7. Actually, the career of the migrant abroad (right panel) remained frozen at the bottom level of the job hierarchy, therefore he/she did not add any credit to his/her social capital abroad: this fact, along with the absence of family contacts and the scarce number of friends in the foreign country explains small increments in utility, as shown in Figure 6 (red line), inducing the return decision. It is worth to notice that here the migrant decided to come back even in absence of any academic call, whereas in the previous example (see Figure 3) he/she remained abroad although in presence of four calls.

The reader should be aware that we have presented the simulations of just two among many possible cases. In order to obtain statistically relevant conclusions about the correlations between the migrant's final decision and the corresponding pair of relevant parameters, i.e. the risk aversion and the initial expectation, we need to run several multi-event simulations that start from many different initial configurations. We will present these results in the next subsection.

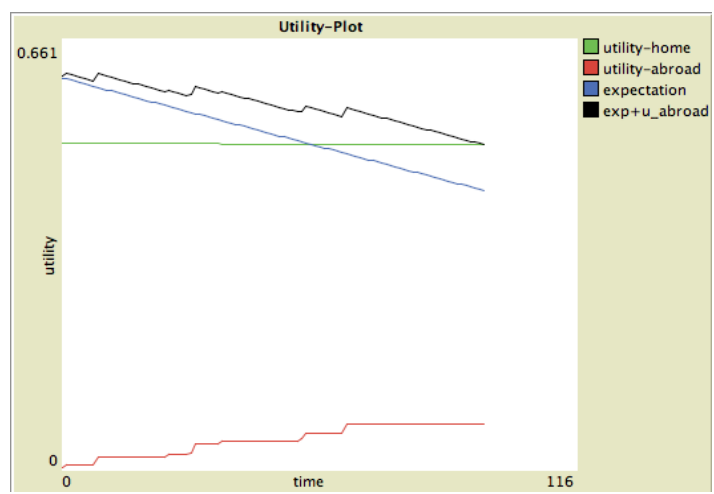

**Figure 6:** Single event dynamics of return migration (return case)

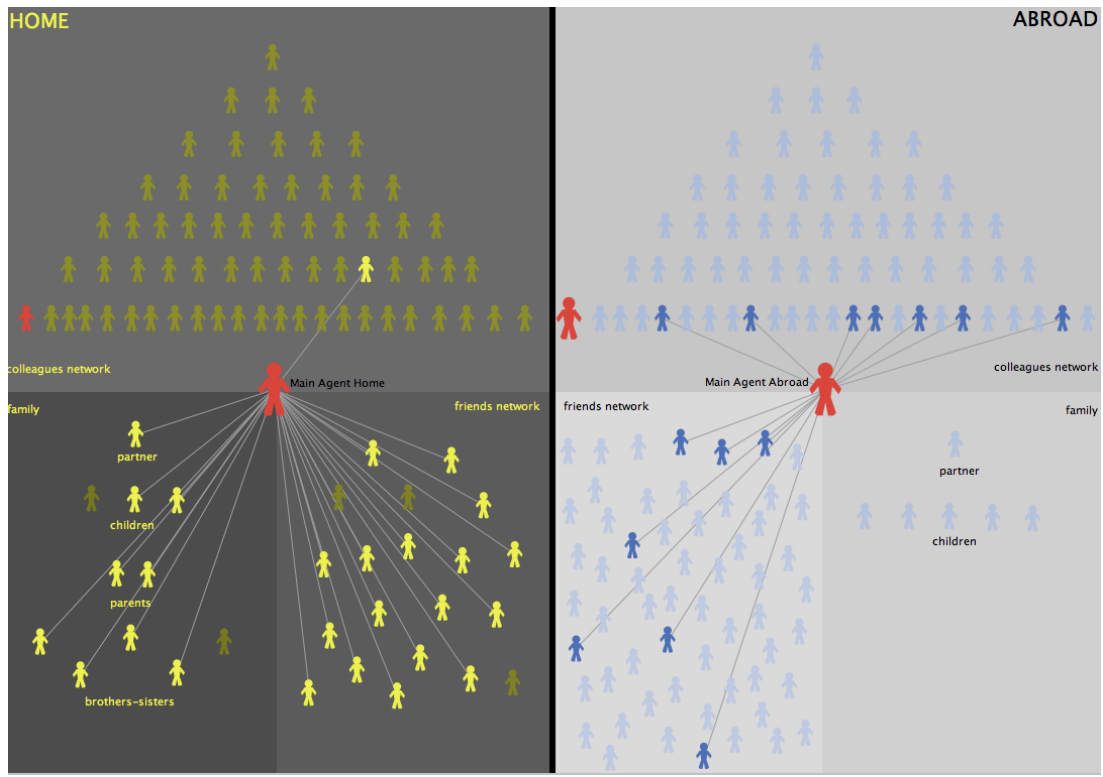

**Figure 7:** Social capitals configuration corresponding to the case plotted in Figure 6

**Multi-event dynamics of return migration**

The multi-event simulation process leads to a full range view of 50000 different events, representing numerous migrants and their final decisions. Four total time intervals $]0,T]$ of 15, 20, 25 and 30 years have been investigated and results appear in Figure 8. Any of the 50000 points drawn in each one of four given "return-phase-diagrams" represents a single simulation in which the initial expectation (x-axis) and the risk-aversion (y-axis) of the migrant are randomly chosen in the interval [0,1]. The colour of the points indicates the final decision of the migrant: if this decision is to return home, the point has been coloured in black; if, instead, the final decision is to remain abroad (for the entire time interval), the point is coloured in red. Notice that dots appear "organized" in vertical strips since the initial expectation (x-axis) is not uniformly extracted in the interval [0,1] but reflects the hierarchical distribution of the relevance for the initial job contact abroad.

Each diagram can be roughly divided in three different regions accordingly with the density of black and red points. The first region, situated in the upper left corner, contains only simulations points corresponding to migrants with a very high risk aversion and a very low initial expectation: these migrants always return to their home country within the time interval, therefore the black colour dominates. We named this region as the "RETURN" zone.

On the contrary, the lower right region of diagrams is exclusively red-coloured. The reason is that the points in this zone represent simulations with very high initial expectation values and very low risk-aversion coefficients for the migrants, who are very well endowed and ready to challenge. This explains why we labelled this region "NO-NETURN" zone. As one could expect, increasing the time interval from 15 to 30 years the RETURN zone tends to expand, whereas the NO-RETURN zone shrinks more and more. In fact, over a broader period, the expectation falls to very small values; therefore even people with a small social capital at home have a very high probability to come back home, especially if meanwhile their social capital abroad has not grown enough.

The essence of this process can be analytically captured by considering the return condition $\sigma < T$, where the return time $\sigma$, recalling eq.(8), can be now rewritten as:

$$\sigma = \frac{u[c^F(\sigma)] - u[c^H(\sigma)]}{k_E a} + \frac{E(0)}{k_E a} < T \ . \qquad (9)$$

This expression allows us to estimate a rough separatrix line between the RETURN and the NO-RETURN zones. Actually, by imposing the parity-equilibrium condition between utilities at home and abroad, i.e. $u[c^F(\sigma)] \approx u[c^H(\sigma)]$, eq.(9) reduces to

$$\frac{E(0)}{k_E a} < T \ \Rightarrow \ a > \frac{E(0)}{k_E T} \ . \qquad (10)$$

Therefore, the equation of the wanted separatrix line is given by

$$a = \frac{E(0)}{k_E T} \qquad (11)$$

which is drawn in yellow in all panels of Figure 8.

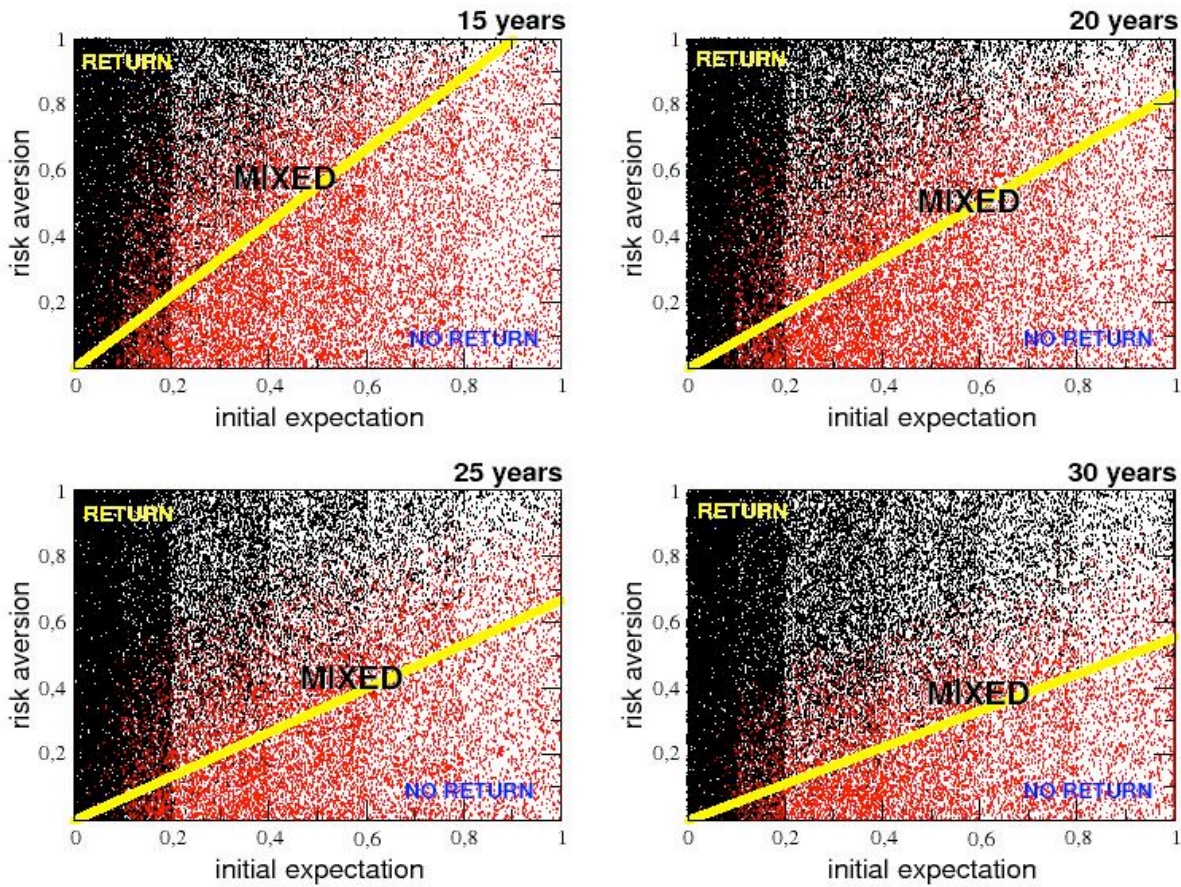

**Figure 8:** Multievent simulation results

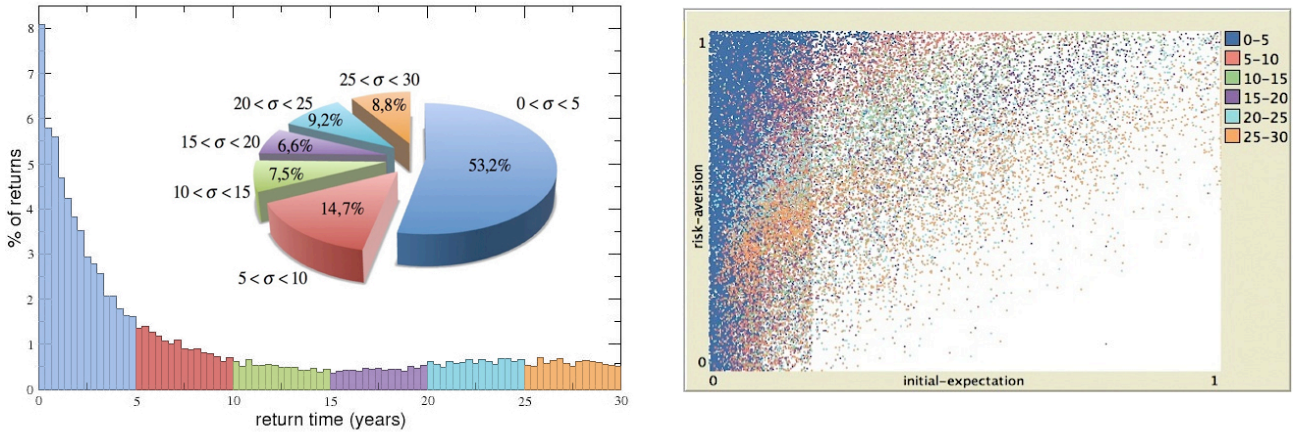

**Figure 9:** Return time distribution

As one can see, in all the four cases considered, this line – whose slope diminishes as the time interval *T* grows – separates the completely red area (the NO-NETURN region) by the remaining part of the phase diagram, where black points (RETURN cases) start to appear. In other words, under the separatrix line the probability for the migrant to came back home vanishes for any initial conditions, while, above the line, the same probability starts to increase, even if in a very complex way (as we will see in detail later) impossible to be described through an analytical approach. Notice also that the linear approximation is not very good for *E(0) < 0.2*.

We remind the reader that in eq.(11) *T* is intended in months, being this the time unit adopted in all simulations, and that we considered $k_E$ = *0.005*: the latter, representing the decay rate of the expectation term, is actually a free parameter of the model, since only an empirical evidence could provide a realistic estimation of its value. At this stage, any numerical choice for $k_E$ (of course within an opportune range), though arbitrary, would provide a coherent scenario, as demonstrated by the good agreement between the prediction of eq.(11) and the simulation results of Figure 8.

Let us now deepen our analysis for return events (black points) shown in Figure 8. In the left panel of Figure 9 we show the frequency distribution of the return times *σ < T* (expressed in percentage) for the simulation with *T = 30* years. In this case we have 36843 events (i.e. about the 74% over the total of 50000) ending with the return of the migrant (black points in the right bottom panel of Figure 8), which are reported in the right panel of Figure 9 with different colours according to different 5-years intervals of return time (see legend). The same intervals, with the corresponding colours, are also indicated in the frequency distribution panel where return times presents a typical power-law decay during the first 20 years, while in the last 10 years the percentage of migrants that decide to came back home slightly increases. In particular, as it can be also appreciated in the cake graph, more than 50% of returns occur within the first 5 years, a period when the initial characterization of the migrant (in terms of risk aversion and initial expectation) plays an important role.

On the other hand, almost 20% of returns are concentrated in time intervals between 20 to 30 years. This timing is particularly far from the initial moment of migration and, therefore, decisions taken in this period are essentially independent of the initial conditions, but relevantly influenced only by the social capital accumulated by the migrant abroad. Such a rationale is confirmed also by looking at the right panel, where the spreading of orange and cyan points (corresponding to the last ten years returns) over a broad region of the diagram, indicates weak correlations with the initial levels of risk aversion and initial expectation.

Let us come back to Figure 8 and focus on the region of the diagrams labelled as "MIXED" zone, immediately above the separatrix line. In this region, black and red points are strongly mixed without any dominance of one colour over the other. This means that, even starting from very similar initial conditions in terms of risk aversion and initial expectation, two different simulations can evolve towards opposite final results, i.e. return or permanence in the foreign country. Such a diverging behaviour depends on two concurrent reasons.

The first one is that, even if the representative points of the two simulations in the phase space were identical, the initial contacts configurations at home and abroad (together with the corresponding social capitals) would be however different. The second one is that the dynamic evolution of events in our model is probabilistic and not deterministic (i.e. there is no "pre-destination"): this means that even the same worker, starting from the same initial configurations of contacts, would not replicate necessarily the same behaviour.

In order to clarify this point, we present a last result coming from 5000 simulations, made by 500 different events, which were performed starting from *exactly the same* initial conditions (in terms of both risk aversion and initial expectation), holding initial social capitals (both at home and abroad) as *constants* and running each event for *T = 15* years. This will discriminate whether migrants with identical social capital endowments eventually assume identical migration decisions or not. In the 3D-graph in Figure 10, the initial expectation and risk-aversion are reported on the *x-y* plane (exactly as in Figure 8).

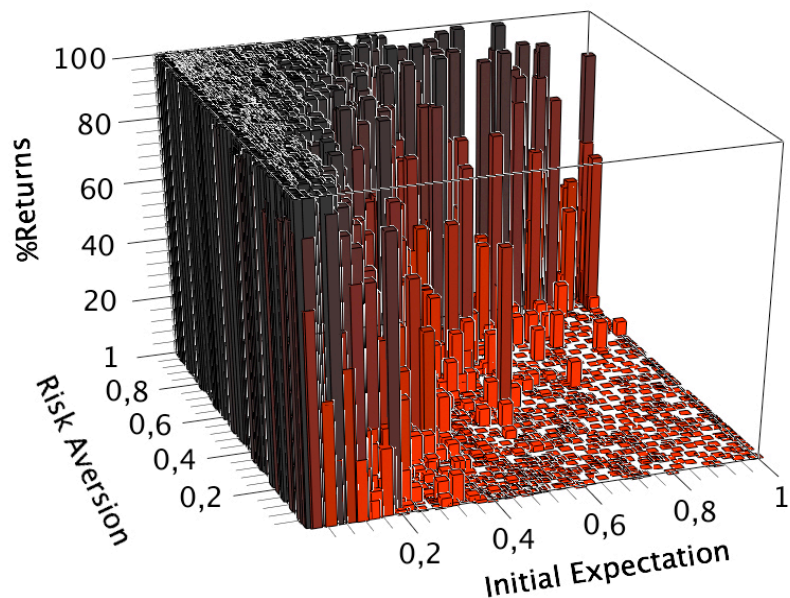

**Figure 10:** Multievent simulation results for identical initial conditions

On the vertical axis, the solid bars' height is proportional to the percentage of events, over the total of 500, where the migrant decided to return for a given couple of the first two variables (it is worth noticing that, looking at Figure 10 from above, it would appear very similar to the corresponding 15-years diagram shown in Figure 8). In other words: the higher the percentage of returns, the taller the vertical bar and the darker its red colour; conversely, the smaller the percentage of returns, the lower the vertical solid, and the brighter its red colour.

It results that only in the two extreme regions corresponding to the RETURN and NO-RETURN zones we have an identical evolution of the 500 events (with either 100% or 0% probability of returns, respectively). Instead, in the MIXED region, the initial conditions are not able to determine uniquely the dynamical evolution of the migration process. This means that it is absolutely not trivial to provide an affordable prediction for migrants' behaviour in this region. Of course a careful calibration of the model's parameters with real datasets might help us to better identify the boundaries of the MIXED region. In any case from this plot one can conclude that, within our approach, the ratio between the risk aversion and the initial expectation $a / E(0)$ is the key factor that discriminates between a NO-RETURN and a RETURN decision: the higher the value of this ratio, the lower the NO-RETURN probability and vice-versa. In particular, as results from Eq.(11) – confirmed by simulations of Figure 8 – for values of this ratio below $1 / k_E T$, the RETURN probability is approximately zero. More precise predictions could be made only after the calibration of model's parameters with real data.

## Conclusions

In this paper we presented a theoretical and a corresponding computational model of return migration after brain drain. Our simulations allow to evaluate the return probability, after a given time interval spent abroad, as function of the initial social capitals of the migrant  (both at home and abroad) and of only two psychological agents' features: the risk aversion and the initial expectation. We found that the final decision of an agent strictly depends on the ratio between these two individual features: if the risk aversion is very high with respect to the initial expectation, the return decision occurs with high probability and vice-versa. In between, there is an intermediate range where large probability fluctuations appear and where the forecast is very difficult since similar or even identical initial conditions lead to different results.

Nonetheless, our model represents an innovative and quantitative tool to model the brain drain phenomenon, which will also allow to extract realistic results once the correct values to the arbitrarily chosen parameters (e.g. the relevance of the different components of migrant's social capital endowment or the decay rate of the expectation) are given. In this respect, it would be extremely useful to have real datasets about migrants in order to calibrate the model, compare our findings with realistic scenarios and validate our predictions. A study in this direction is in progress and will be presented in a forthcoming paper.

## Notes

NetLogo Model can be retrieved from OpenABM: http://www.openabm.org/model/3420/version/1